\documentclass[twocolumn,prl,preprintnumbers,amsmath,amssymb,letter]{revtex4-1}
\usepackage{graphicx}   
\usepackage{longtable}
\usepackage{graphicx}   

\begin{document}

\title{Generation of
a Complete Set of Additive Shape Invariant Potentials from an Euler Equation}
\author{Jonathan Bougie}
 \email{$\!$jbougie@luc.edu, $\dag$agangop@luc.edu, $\ddag$jmallow@luc.edu }   
 \affiliation{Loyola University Chicago, Department of Physics, Chicago,
IL 60660}
\author{Asim Gangopadhyaya$^{\dag}$}
 \affiliation{Loyola University Chicago, Department of Physics, Chicago,
IL 60660}
\author{Jeffry V. Mallow$^{\ddag}$}
 \affiliation{Loyola University Chicago, Department of Physics, Chicago,
IL 60660}
\date{\today}

\begin{abstract}
In supersymmetric quantum mechanics, shape invariance is a sufficient condition for solvability. We show that all conventional additive shape invariant superpotentials that are independent of $\hbar$ obey two partial differential equations. One of these is equivalent to the one-dimensional Euler equation expressing momentum conservation for inviscid fluid flow, and it is closed by the other. We solve these equations, generate the set of all conventional shape invariant superpotentials, and show that there are no others in this category. We then develop an algorithm for generating all additive shape invariant superpotentials including those that depend on $\hbar$ explicitly.
\end{abstract}
\maketitle
\noindent
{PACS numbers: 03.65.-w, 47.10.-g, 11.30.Pb}

\vspace*{0.1in}

Supersymmetric quantum mechanics (SUSYQM) \cite{Witten,Cooper1,Cooper} is a generalization of the ladder operator formalism usually attributed to Dirac  \cite{Dirac}. It makes use of first order differential operators
$A^{\pm} \equiv \mp\, \hbar \, \frac{d } {dx} +W(x,a)$, where the superpotential $W(x,a)$ is a real function of $x$, and $a$ is a parameter. We define two partner hamiltonians
\begin{eqnarray} H_{\mp} &=& A^{\pm}\, A^{\mp}
= -\,\hbar^2\,{{d^2} \over {dx^2}}+ V_{\mp}(x,a)~,
\label{Schrodinger}
\end{eqnarray}
where partner potentials $V_{\pm}(x,a)$ are related to the superpotential by $V_{\pm}(x,a)=W^2(x,a)\pm \hbar \, {{dW(x,a)} \over {dx}}$. Partner hamiltonians
 \cite{units} have the same energy eigenvalues (except for the ground state); i.e., $E_{n+1}^{(-)}=E_{n}^{(+)}$ and $E_{0}^{(-)}=0$.  Eigenfunctions of $H_{\mp}$ are related by: $\psi^{(+)}_{n-1} \propto A^-(x,a) \psi^{(-)}_n$ and $A^+(x,a) \psi^{(+)}_{n-1} \propto \psi^{(-)}_{n}$. Thus, if the eigenvalues and the eigenfunctions of $H_{-}$ are
known {\it a priori}, they are automatically determined for $H_{+}$ as well.

If the partner potentials $V_{\pm}(x,a)$ obey the ``shape invariance" condition \cite{Infeld, gendenshtein},
\begin{eqnarray}
V_+(x,a_0)+
g(a_0)~=~V_-(x,a_1)+  g(a_1)~, %
\label{SIC-Potential}
\end{eqnarray}
then the spectrum for either Hamiltonian can be derived without reference to its partner.  This is due to the existence of an underlying potential algebra  \cite{balantekin,asim1,asim2}.

Shape invariant partners have the same form except for the value of the parameter $a_i$, where $a_1$ is a function of $a_0$; i.e.,  $a_1 = f(a_0)$.
The energy eigenvalues of $H_-(x,a_0)$ are given by
$E^{(-)}_n(a_0) = g(a_n) -g(a_0)$, where $a_n \equiv f^n(a_0)$  indicates $f$ applied $n$-times to $a_0$
 \cite{Cooper}.
If the parameters differ only by an additive constant: $a_{i+1} = a_{i}+\hbar$, the potentials are called ``additive" or ``translational" shape invariant. All exactly solvable potentials discovered thus far that are expressible in terms of known functions are additive shape invariant \cite{Cooper,Dutt}. Several groups found these potentials by imposing various ansatzes  \cite{CGK,asim2,asim3, Quesne}.

Important correspondences exist between quantum mechanics and fluid mechanics \cite{Curtright}. SUSYQM is well known to have a deep connection with the KdV equation  \cite{KdV, sukumar1, sukumar2, wang, kwong}, a nonlinear equation that describes waves in shallow water.
We now prove that every additive shape invariant superpotential that does not depend on $\hbar$ explicitly corresponds to a solution of the Euler equation expressing momentum conservation for inviscid fluid flow in one spatial dimension. We use this correspondence to find a systematic
method which (1) yields all known such superpotentials for SUSYQM and (2) shows that no others exist.  We then extend this method to general additive shape invariant superpotentials including those that depend on $\hbar$ explicitly  \cite{Quesne}.

Writing Eq.~(\ref{SIC-Potential}) in terms of the superpotential yields
\begin{eqnarray}
W^2(x,a_0) &+& \hbar \, \frac{d W(x,a_0)}{d x} + g(a_0)\nonumber \\&=&W^2(x,a_1)
- \hbar \,
\frac{d W(x,a_1)}{d x} + g(a_1) ~~.%
\label{SIC-SuperPotential}\end{eqnarray}

\vspace{-0.05in}
\noindent
Eq.~(\ref{SIC-SuperPotential}) is a difference-differential equation relating the square of the superpotential $W$ and its spatial derivative computed at two different  parameter values: $(x,a_0\equiv a)$ and $(x,a_1
\equiv a+ \hbar)$. This equation must hold for any value of $\hbar$.  At this point we consider only superpotentials that do not depend explicitly on $\hbar$, but only depend on $\hbar$ through the parameter $a$; we will call this class  ``conventional''. We will later consider general superpotentials that may depend on $\hbar$ explicitly. We show that the shape invariance condition (Eq.~\ref{SIC-Potential}) can be
expressed as a local non-linear partial differential equation; i.e., all terms can be computed
at the same point $(x,a)$. This will provide a  systematic method for finding superpotentials.

Since  Eq.~(\ref{SIC-SuperPotential}) must hold for any value of $\hbar$, if we expand in powers of $\hbar$, the coefficient of each power must separately vanish. Provided that $W$ does not depend explicitly on $\hbar$, this expansion yields
\begin{eqnarray}%
\!\!\!\!\!\!\!\!{\cal O}(\hbar)&\Rightarrow& W \, \frac{\partial W}{\partial a} - \frac{\partial W}{\partial x} + \frac12 \, \frac{d g(a)}{d a} = 0~,
\label{PDE1}\\
\!\!\!\!\!\!{\cal O}(\hbar^2)&\Rightarrow& \frac{\partial }{\partial a}\left( W \, \frac{\partial W}{\partial a} - \frac{\partial W}{\partial x} + \frac12 \, \frac{d g(a)}{d a} \right)= 0~,
\label{PDE2}\\
\!\!\!\!\!\!{\cal O}(\hbar^n)&\Rightarrow& \frac{\partial^{n}}{\partial a^{n-1}\partial x} ~W(x,a)= 0~, ~~~~~n\geq 3 ~.
\label{PDE3} \end{eqnarray}
Thus, all conventional additive shape invariant superpotentials are solutions of Eqns.~(\ref{PDE1}-\ref{PDE3}).  Although this represents an infinite set, note that if equations at ${\cal O}(\hbar)$ and ${\cal O}(\hbar^3)$ are satisfied, all others automatically follow.

Replacing $W$ by $-u$, $x$ by $t$, and $a$ by $x$
in Eq.~(\ref{PDE1}), we obtain:
\begin{eqnarray}%
u(x,t) \,  \frac{\partial  }{\partial
x} u(x,t)+ \frac{\partial u(x,t) }{\partial t}  = -\frac12
\frac{d g(x)}{d x}~.
\end{eqnarray}
This is equivalent to the equation for inviscid fluid flow in the absence of external forces on the body of the fluid:
\begin{equation}\frac{\partial {\bf u}\left({\bf x},t\right)}{\partial t}+u\left({\bf x},t\right)\cdot\nabla {\bf u}\left({\bf x},t\right) = -\frac{\nabla p\left({\bf x},t\right)}{\rho\left({\bf x},t\right)}\label{Euler}\end{equation}
 in one spatial dimension with the correspondence $\frac{1}{\rho}\frac{d p}{d x} = \frac 12 \frac{d g}{d x}$,
where ${\bf u}$ is the fluid velocity at
location ${\bf x}$ and time $t$, $p$ is the pressure, and $\rho$ is the local fluid density.
Equation~(\ref{Euler}) is one of the fundamental laws of fluid dynamics, and was first obtained by Euler
in 1755  \cite{Euler}.  Thus, all conventional shape invariant superpotentials form a set of solutions to the one-dimensional Euler equation.

Note that Eq.~(\ref{Euler}) is not closed as written.  In fluid dynamics this equation is generally supplemented by the continuity equation expressing conservation of mass, along with an equation of state and/or the energy equation and boundary conditions. These additional constraints do not apply in SUSYQM.  Instead, Eq.~(\ref{PDE3}) supplies the additional constraint.

Equation~(\ref{PDE3}) is satisfied for all $n\geq 3$ as long as
\begin{eqnarray}
\frac{\partial^{3}}{\partial a^{2}\partial x} ~W(x,a)= 0.\label{Eq1}
\end{eqnarray}
The general solution to Eq.~ (\ref{Eq1}) is
\begin{eqnarray}
W(x,a)=a\cdot {X}_1(x)+{X}_2(x)+u(a)~.\label{GeneralSolution}
\end{eqnarray}
Substituting this into Eq.~(\ref{PDE1}), and collecting and labeling terms based on their dependence on ${X}_{1}$ and ${X}_{2}$, we obtain
\vspace*{-0.25in}
\begin{widetext}
\begin{eqnarray}
\underbrace{{X}_1\,{X}_2}_{\mbox{Term\#1}} +
\underbrace{\left(- \frac{d {X}_2}{d x}\right)}_{\mbox{Term\#2}}
+
\underbrace{a\, {X}_1^2}_{\mbox{Term\#3}}+
\underbrace{\left( - a \frac{d {X}_1}{d x}\right)}_{\mbox{Term\#4}}+
\underbrace{\frac{d u}{d a}\,{X}_2 }_{\mbox{Term\#5}}+
\underbrace{\left(u+a\frac{d u}{d a}\right){X}_1}_{\mbox{Term\#6}}=
H(a)~,\label{Eqterms}
\end{eqnarray}
\end{widetext}
where $H(a)\equiv - u\, \frac{d u}{d a} -\frac12~\frac{d g}{d a}$.
To find all possible solutions, we begin by considering special cases of Eq.~(\ref{Eqterms}) where one or more of the terms ${X}_1(x)$, ${X}_2(x)$, or $u$ is zero.  After considering these cases, we will show that all solutions can be reduced to one of these cases.
In our nomenclature, lower case Greek letters denote $a$- and $x$-independent constants.

{\bf Case 1: ${X}_{2}$ and $u$ are not constants, ${X}_{1}$ is constant.}  In this case, let ${X}_{1}=\mu$. Then  $W=\mu a + u(a) + {X}_{2}(x)$.  If we define $\tilde{u}\equiv u(a)+\mu a$, we get $W=\tilde{u}+{X}_{2}$.  So this case is equivalent to ${X}_{1}=0$.   Then terms 1, 3, 4, and 6 each becomes zero, and Eq.~(\ref{Eqterms}) becomes $-\frac{d {X}_{2}}{dx}+\frac{du}{da}{X}_{2}= H(a)$.    Since ${X}_{2}$ must be  independent of $a$, $\frac{du}{da}$ and $H(a)$ must be constants. This yields $u=\alpha a + \beta$ and $-\frac{d {X}_{2}}{dx}+\alpha{X}_{2}= \theta$.  The solution is ${X}_{2}(x)=\frac{\theta}{\alpha}+\eta\, e^{\alpha x}$.  Therefore, $W=\alpha a +\beta+\frac{\theta}{\alpha}+\eta\, e^{\alpha x}$. Defining $\alpha = -1$, we obtain $W=A-B e^{-x}$, where $A\equiv \beta-a-\theta$.  This is the Morse superpotential.

{\bf Case 2: ${X}_{1}$ and $u$ are not constants, ${X}_{2}$ is constant.}  Following a similar procedure, this case is equivalent to ${X}_{2}=0$.  Depending on the values of constants of integration, this equation yields the Rosen-Morse I, Rosen-Morse II, Eckart, and Coulomb superpotentials.

{\bf Case 3: ${X}_{1}$ and ${X}_{2}$ are not constants, $u=\mu a + \nu$.} We define $\tilde{X}_{1} \equiv {X}_{1}+\mu$ and $\tilde{X}_{2} \equiv {X}_{2}+\nu$, making this case equivalent to $u=0$.  Depending on the constants of integration, this yields the Scarf I, Scarf II, 3-D oscillator, and generalized P\"{o}schl-Teller superpotentials.

{\bf Case 4: ${X}_{2}$ is not constant, ${X}_{1}$ and $u$ are constant.}
If ${X}_{1}\neq 0$ we get Morse,  and ${X}_{1}=0$ generates the one-dimensional harmonic oscillator.

{\bf Case 5: ${X}_{1}$ is not constant, ${X}_{2}$ and $u$ are constant.}
This yields special cases of Scarf I and Scarf II, and the centrifugal term of the Coulomb and 3-D oscillator.

{\bf Case 6: ${X}_{1}$ is constant, ${X}_{2}$ is constant.}
In this case, the superpotential has no $x$-dependence, regardless of the value of $u$.  This is a trivial solution corresponding to a flat potential, and we disregard it.

These special cases generate all known conventional additive shape-invariant superpotentials \cite{Dutt,Cooper}, as shown in Table \ref{Table1}.

Now that we have considered these special cases, we can systematically obtain all possible solutions.  $H(a)$ is independent of $x$.  Therefore, when any solution is substituted into Eq.~(\ref{Eqterms}), it must yield an $x$-independent sum of terms 1-6.  There are many ways in which these terms could add to a sum independent of $x$. We begin with the simplest possibility, in which each term is individually independent of $x$.  In this case, term 3 states that ${X}_{1}$ must be a constant, independent of $x$.  In addition, term 1 dictates ${X}_{1} {X}_{2}$ must be constant as well.  These two statements can only be true if ${X}_{2}$ and ${X}_{1}$ are constant separately; this reduces to the trivial solution of case 6.

Therefore, assuming that each term is separately independent of $x$ yields only the trivial solution.  However, there is also the possibility that some of the terms depend on $x$, but when added to other terms, the $x$-dependence cancels to yield a sum that is independent of $x$.   If a group of $n$-terms taken together produces an $x$-independent sum, and if no smaller subset of these terms add up to a sum independent of $x$, we call this group ``irreducibly independent of $x$."

\begin{table} [htb]
\begin{center}
\begin{tabular}{||l|l|l||}
\hline Name  &  superpotential   & Special \\
&  & Cases \\ \hline
Harmonic Oscillator &  $\frac12 \omega x$ & $X_1=u=0$  \\
  \hline
Coulomb   &  $\frac{e^2}{2(\ell+1)} - \frac{\ell+1}{r}$& $X_2=0$  \\
    \hline
3-D oscillator  & $\frac12 \omega r - \frac{\ell+1}{r}$ &  $u=0$\\
 \hline
Morse &$A-Be^{-x}$ &  $X_1=0$ \\
 \hline
Rosen-Morse I &$-A\cot x - \frac{B}{A}$ & $X_2=0$  \\
 \hline
Rosen-Morse II &$A\tanh x + \frac{B}{A}$ & $X_2=0$   \\
 \hline
Eckart &$-A\coth x + \frac{B}{A}$ & $X_2=0$  \\
 \hline
{Scarf I} &$ A\tan x - B {\rm sec}\,x$ & $u=0$\\
 \hline
{Scarf II}  &$A\tanh x + B {\rm sech}\,x$& $u=0$\\
 \hline
Gen. P\"oschl-Teller &$A\coth x- B {\rm cosech}\,x $ & $u=0$ \\
 \hline
\end{tabular}
\caption{The complete family of conventional additive shape-invariant superpotentials.}\label{Table1}
\end{center}
\end{table}

If, for example, term 2 depends on $x$ and term 5 depends on $x$, but the sum of these two terms is $x$-independent, then we consider the set of terms $\{2,5\}$ to be a two-term set that is irreducibly independent of $x$.  Let us investigate this example further.

In this case, $-\frac{d{X}_{2}}{dx}+\frac{du}{da}{X}_{2}$ is independent of $x$.  However, ${X}_{2}$ and $\frac{d{X}_{2}}{dx}$ must each depend on $x$, or this would be reducible.  Since term 2 does not depend on $a$, $\frac{du}{da}$ must be constant for the $x$-dependence of terms 2 and 5 to cancel.  So $u=\delta_{1}a+\delta_{2}$. Substituting this into terms 2 and 5 yields $-\frac{d{X}_{2}}{dx}+\delta_{1}{X}_{2}=\delta_{3}$.  The solution is ${X}_{2}=\frac{\delta_{3}}{\delta_{1}}- \delta_{4}e^{\delta_{1} x}$.
For this solution to work, the sum of the remaining terms 1, 3, 4, and 6 must also be independent of $x$.  We first ask if this could be true by making all of the remaining terms each independent of $x$.  This is only possible if ${X}_{1}=0$.  Thus, the combination where terms 1, 3, 4, and 6 are each individually independent of $x$ and $\{2,5\}$ is an irreducibly independent set is an example of Case 1 above (since ${X}_{1}=0$) and yields the Morse superpotential.

We continue in this manner, checking whether each two-term irreducible set yields solutions when combined with the remaining terms each independent of $x$ as in the example above.  In each case, we find either that the equation reduces to one of the special cases examined earlier, or that no solution is allowed (for instance, term 1 and term 3 cannot be irreducibly independent of $x$ since one is independent of $a$ and the other is linear in $a$).

Once these combinations are exhausted, we consider combinations of two-term irreducible sets with other two-term irreducible sets as well as single-term constants.  Then we examine three-term irreducible sets, all the way up to the full six-term equation.

As a final example, we check whether there are any solutions for the full six-term irreducible set.
We note that the first two terms are independent of $a$, while terms 3 and 4 are linear in $a$.  We do not know {\it a priori} the functional form of $u$.  However, we do know that any $x$-dependence in terms 1 and 2 cannot be canceled by terms 3 and 4, since terms linear in $a$ cannot cancel terms independent of $a$.  For an irreducible set, the sum of the first two terms must have an $x$-dependence that is canceled by $a$-independent terms from $u$ and $\frac{du}{da}$ in terms 5 and 6, and terms 3 and 4 must have an $x$-dependence that is cancelled by terms linear in $a$.  Since term 5 contains $\frac{du}{da}$, it could include terms independent of $a$, terms linear in $a$, and/or other forms of $a$-dependence.  However, it cannot fully cancel the $x$-dependence of the first four terms or the set will be reducible.

We conclude that the only way for the solution to be irreducible is if $u+a\frac{du}{da}=\xi \frac{du}{da}+\mu a + \nu$ for constants $\mu$, $\nu$, and $\xi$.  This gives $u=\frac{\mu a}{2}+ \left(\xi + \frac{\mu \nu}{2}\right) + \frac{\gamma}{a-\nu}$.  By collecting terms of the same power in $a$, we find that the terms proportional to $\frac{1}{(a-\nu)^2}$ mandate that $\gamma\left({X}_{2}+\nu{X}_{1}\right)$ is a constant.  This leaves only two possibilities.  First, $\gamma=0$, in which case $u$ depends only linearly on $a$, and this reduces to Case 3 above.  Otherwise, ${X}_{2}+\nu {X}{_1}$ must be a constant.  In this case, ${X}_{1}$ differs from ${X}_{2}$ by only a multiplicative constant; by shifting the zero of $a$, we can absorb ${X}_{2}$ into ${X}_{1}$, and this reduces to Case 2 above.  Thus, any solution for the full set of terms can be reduced to one of the special cases.

By examining all possible combinations of terms, we have found that no new solutions are admitted by any combinations that are not included as one of our special cases.  Thus, we have found all known additive shape-invariant superpotentials that do not depend explicitly on $\hbar$, and have proven that no more can exist.

However, a new family of ``extended" shape-invariant potentials was recently discovered by Quesne \cite{Quesne}, and expanded elsewhere \cite{odake}.  These potentials are generated from our system by generalizing our formalism to include superpotentials that contain $\hbar$ explicitly. In this case, we expand the superpotential $W$ in powers of $\hbar$:
\begin{eqnarray}
    W(x, a, h) = \sum_{n=0}^\infty \hbar^n W_n(x,a)~. \label{W-hbar}
\end{eqnarray}
Substituting Eq.~(\ref{W-hbar}) in Eq. (\ref{SIC-SuperPotential}), significant algebraic manipulation yields
\begin{widetext}
\begin{eqnarray}
\sum_{n=1}^\infty \hbar^n \left[\sum_{k=0}^n W_k\, W_{n-k} + \frac{\partial W_{n-1}}{\partial x}
-\sum_{s=0}^n \sum_{k=0}^s \frac{1}{(n-s)!}
\frac{\partial^{n-s}}{\partial a^{n-s}} W_k\, W_{s-k}
+\sum_{k=0}^{n-1}\frac{1}{(k-1)!}
\frac{\partial^{k+1}}{\partial a^k \,\partial x} \, W_{n-k-1}- \left( \frac1{n!}\,\frac{\partial^{n} g}{\partial a^n}   \right)\right]=0.\nonumber
\end{eqnarray}
As this must hold for any value of $\hbar$, the following equation must hold separately for each positive integer value of $n$:
\begin{eqnarray}
\sum_{k=0}^n W_k\, W_{n-k} + \frac{\partial W_{n-1}}{\partial x}
-\sum_{s=0}^n \sum_{k=0}^s \frac{1}{(n-s)!}
\frac{\partial^{n-s}}{\partial a^{n-s}} W_k\, W_{s-k}
+\sum_{k=1}^{n} \frac{1}{(k-1)!}
\frac{\partial^{k}}{\partial a^{k-1} \,\partial x} \, W_{n-k}- \left( \frac1{n!}\,\frac{\partial^{n} g}{\partial a^n}   \right)=0.
\label{higherorders}\end{eqnarray}
\end{widetext}
For $n=1$, we obtain
\begin{eqnarray}
2\frac{\partial W_{0}}{\partial x}-\frac{\partial }{\partial a} \left(W_{0}^2+g  \right) = 0,
\end{eqnarray}
yielding $2\frac{\partial^k W_{0}}{\partial a^k\partial x}= \frac{\partial^k }{\partial a^k} \left(W_{0}^2+g  \right)~{\rm for}~k\ge 1$. We have shown that all conventional superpotentials $W=W_0$ are solutions of this equation.  Higher order terms can be generated from applying  Eq.~(\ref{higherorders}) for all $n>1$.

As an example, we choose the 3-D oscillator solution: $W_{0} = \frac12 \omega x - \frac ax$.
For $n=2$, the expansion yields
\begin{eqnarray}
\frac{\partial W_{1}}{\partial x}-\frac{\partial }{\partial a} \left(W_{0}W_{1}  \right) 
= 0~,\nonumber
\end{eqnarray}
and for $n=3$, we obtain
\begin{eqnarray}
\frac{\partial W_{2}}{\partial x}-\frac{\partial \left(2W_{0}W_{2}+W_{1}^2  \right)}{\partial a}
- \frac 12 \frac{\partial^2 W_{0}W_{1}}{\partial a^2}
+ \frac 23 \frac{\partial^3 W_0}{\partial a^2\partial x}  = 0.\nonumber
\end{eqnarray}
These two coupled equations are solved by $W_{1}=0$ and
$W_2 = (4 x \omega)/(2 a+x^2 \omega)^2$.
The next order equations are solved by $W_{3}=0$ and $W_4 = (4 x \omega)/(2 a+x^2 \omega)^4$.
Generalizing these, we get  $$W_0=\frac12 \omega x - \frac ax;~W_{2n+1}=0;~W_{2n} = (4 x \omega)/(2 a+x^2 \omega)^{2n},$$
yielding a sum that converges to
$$W(x,a,\hbar) =  \frac12 \omega x - \frac ax + \left(\frac{2\omega x \hbar }{\omega x^2+2a-\hbar} - \frac{2\omega x \hbar }{\omega x^2+2a+\hbar}\right).$$
With the identification $a=(\ell+1)\hbar$, and $\hbar = 1$,
\begin{eqnarray}
W &\rightarrow& \frac{\omega x}2  - \frac {\ell+\!1}x + \left(\frac{2\omega x }{\omega x^2+2\ell+\!1} - \frac{2\omega x }{\omega x^2+2\ell+\!3}\right).\nonumber\end{eqnarray}
This is the extended superpotential found by Quesne
\cite{Quesne}.

We have thus obtained a system of partial differential equations that must be satisfied for all shape-invariant superpotentials.  For conventional cases that do not depend on $\hbar$, we have shown that the shape invariance condition is equivalent to an Euler equation expressing momentum conservation for fluids and an equation of constraint. For extended cases in which the superpotential depends explicitly on $\hbar$, we  developed an algorithm that is satisfied by all additive shape invariant superpotentials.

\begin{acknowledgements}

We thank the referees for their invaluable suggestions for improving the manuscript. In particular the suggestion to include extended superpotentials added to the completeness of the manuscript.

This research was supported by an award from Research Corporation for Science Advancement. 
\end{acknowledgements}


\begin{thebibliography}{99}
\bibitem{units}We are using units for particle mass such that $2m=1$ in the Schroedinger equation.
\end{thebibliography}
\end{document}